\UseRawInputEncoding

\documentclass[aps,prr,twocolumn,superscriptaddress,showpacs,amsmath,amssymb, floatfix]{revtex4-2}

\usepackage{graphicx,caption}
\usepackage{latexsym}
\usepackage{amsmath}
\usepackage{amssymb}
\usepackage{amsfonts}
\usepackage{color}
\usepackage{bm}
\usepackage{verbatim}
\usepackage{pagecolor,lipsum}
\usepackage{float}
\usepackage{hyperref}
\usepackage[normalem]{ulem}
\hypersetup{colorlinks=true,urlcolor=blue,linkcolor=blue,citecolor=blue}
\usepackage{comment}
\usepackage{soul,xcolor}
\usepackage{mhchem}
\usepackage{stfloats}
\usepackage{array}

\usepackage{xcolor}

\definecolor{MS-color}{RGB}{128,0,128}

\setstcolor{red}

\begin{document}
\title[Triplet Supercurrents in half-metallic ferromagnets]{Triplet supercurrents in lateral Josephson junctions with a half-metallic ferromagnet}

\author{Yao Junxiang}
\email{yao@Physics.LeidenUniv.nl}
\affiliation{Huygens-Kamerlingh Onnes Laboratory, Leiden Institute of Physics, Leiden University. P.O. Box 9504, 2300 RA Leiden, The Netherlands.}

\author{Remko Fermin}
\affiliation{Department of Materials Science $\&$ Metallurgy, University of Cambridge, 27 Charles Babbage Road, Cambridge CB3 0FS, United Kingdom.}

\author{Mariona Cabero}
\affiliation{ICTS - Centro Nacional de Microscopía Electrónica, Universidad Complutense de Madrid, 28040 Madrid, Spain}

\author{Kaveh Lahabi}
\affiliation{Huygens-Kamerlingh Onnes Laboratory, Leiden Institute of Physics, Leiden University. P.O. Box 9504, 2300 RA Leiden, The Netherlands.}

\author{Jan Aarts}
\email{aarts@physics.leidenuniv.nl}
\affiliation{Huygens-Kamerlingh Onnes Laboratory, Leiden Institute of Physics, Leiden University. P.O. Box 9504, 2300 RA Leiden, The Netherlands.}

\date{\today}
%




\raggedbottom

\begin{abstract}
In the area of superconducting spintronics, spin triplet supercurrents in half-metallic ferromagnets (HMFs) could yield dissipationless spin transport over large distances, and high current density. Promising among the HMFs is the perovskite oxide La$_{0.7}$Sr$_{0.3}$MnO$_3$ (LSMO), and recent studies in combination with the high-T$_c$ superconductor YBa$_2$Cu$_3$O$_7$, or the conventional superconductor NbTi, showed long range effects. Here we focus on two issues that as yet received less attention: the value of the critical current in the HMF in the limit of a very small electrode distance (20~nm), and the nature of the spin triplet generator. We use lateral junctions shaped as bar, square, and disk, and find high supercurrent densities, of order 10$^{11}$~A/m$^2$, pointing to an efficient triplet generation mechanism. This is surprising in the sense that no magnetic inhomogeneity is purposely built in, as is done in conventional metal triplet junctions. Furthermore, from the magnetic field dependence of the critical current interference patterns we find a uniform supercurrent distribution in bar-shaped devices, but one more constricted to the rim in disk devices, which is an expected consequence of the geometry. We also analyze the temperature dependence of the critical current and find the quadratic dependence that was predicted in the limit of small junction lengths. From studying the NbTi/LSMO interface with scanning electron transmission microscopy, we conclude that the magnetic inhomogeneity required for triplet generation resides in the LSMO layer adjacent to the interface.
\end{abstract}

\maketitle


\section{Introduction}\label{sec1}

Arguably the most promising ferromagnetic (F) materials to combine with a superconductor (S) for superspintronics applications~\cite{2015Eschrig,2015Linder,Robinson2021} are the half-metallic ferromagnets (HMF), such as CrO$_2$ and the metallic oxide perovskites La$_{0.7}$X$_{0.3}$MnO$_3$ (X = Sr (LSMO); or Ca (LCMO)). In both types of HMF materials, long-range triplet (LRT) proximity effects were found in lateral Josephson junctions, showing the existence of supercurrents over extremely long length scales~\cite{2006Keizer,Anwar2010,Singh2016,2024Yao}: in CrO$_2$ over a distance of almost 0.5~{\textmu}m~\cite{2006Keizer,Singh2016} and in LSMO even bridging 1~{\textmu}m~\cite{2022Sanchez,2024Yao}. Not well understood, however, is what magnitude of the supercurrent can be expected. This touches upon the question what the generator of spin triplets is in the LSMO-based experiments. In conventional S/F systems, a spin-active interface is engineered at which spin mixing and spin rotation processes conspire to the formation of equal-spin triplet correlations~\cite{2007Houzet,2008Eschrig}. Stack of magnets with non-collinear magnetizations are generally used for the magnetic inhomogeneity~\cite{2010Khaire,2010Robinson,2016Singh}. More recently, magnetic vortices~\cite{Kalenkov2011,Silaev2009,2017Lahabi,2022Fermin}, domain walls~\cite{Bergeret2001a,Fominov2007,Kalcheim2011,Aikebaier2019} and spin-orbit coupling~\cite{Niu2012,Bergeret2013_New,Bergeret2014a,Alidoust2015,jacobsen2015,Bujnowski2019,Eskilt2019,Silaev2020} were also identified as triplet generators. The results on CrO$_2$ without engineered magnetic non-collinearity are hypothesized to result from intrinsic strain-induced magnetic inhomogeneity or grain boundary disorder~\cite{Anwar2011}, but there are no engineered or other established sources of spin triplet correlations in the recent experiments on LSMO junctions with either the high temperature superconductor YBa$_2$Cu$_3$O$_7$~\cite{2022Sanchez}, or the conventional superconductor NbTi~\cite{2024Yao}.\\

\begin{figure}[ht]
\centering
\includegraphics[width=6cm]{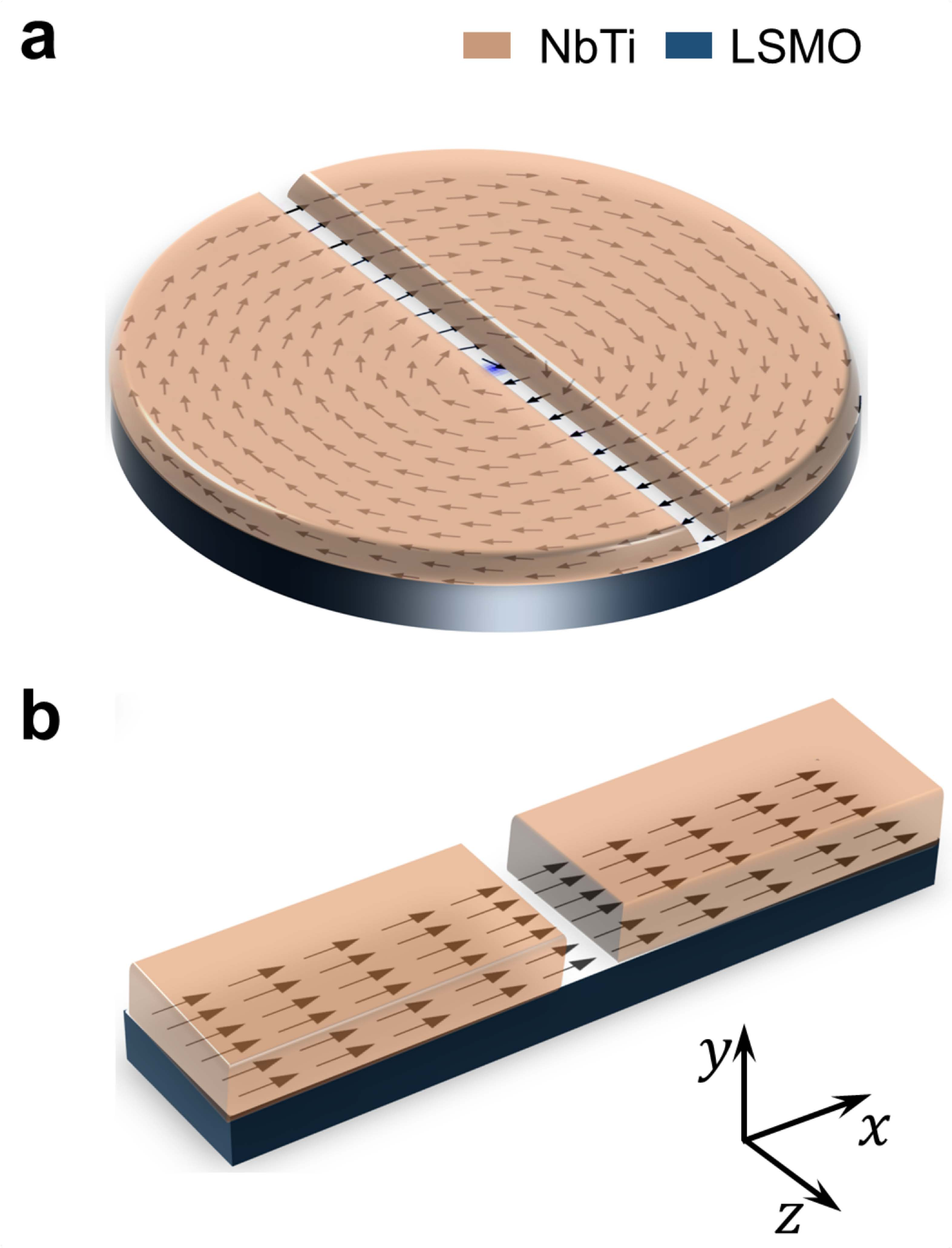}
\caption{Schematic illustration of LSMO/NbTi junctions of different geometry. (a) A disk-shaped junction consisting of NbTi (orange) and LSMO (blue). Due to the shape anisotropy, a stable magnetic vortex forms in the LSMO layer, as indicated. (b) A bar-shaped junction with a uniform magnetic texture. The $x(z)$-axis is perpendicular (parallel) to the trench (both in the sample plane). The out-of-plane direction is the $y$-axis.}\label{disk-fig1}
\end{figure}

Earlier we studied lateral junctions of LSMO/NbTi, fabricated by electron beam lithography, and with a large distance between the contacts, of minimally 500~nm, which made them diffusive junctions. Here, we go to the opposite regime and fabricate junctions with a gap-between-electrodes of around 20~nm, the smallest we can make by a focused ion beam. We study different geometries and compare disk junctions, where spin texture actually can generate triplets~\cite{2022Fermin}, with square and bar-shaped junctions without engineered magnetic inhomogeneity (see Fig.\ref{disk-fig1}). We find very high critical currents in our devices. Irrespective of the device geometry, the value at 4.2~K is around 10$^9$~A/m$^2$, with the disk device reaching 10$^{11}$~A/m$^2$ at 1.5~K. We also find a current distribution that depends on the geometry. Analysis of the critical current interference patterns upon application of a perpendicular magnetic field shows that in the disk, the supercurrent is highly constricted to its rims, similar to results we obtained on disk junctions made of Nb/Co~\cite{2022Fermin}. For the bars we find a much more homogeneous distribution over its cross-section. The results show that the triplet generator mechanism has to be found in the LSMO/NbTi interface, although the spin texture of the disk has a decided influence. From electron energy loss spectroscopy (EELS), performed with a scanning transmission electron microscope (STEM), we find indications of oxygen inhomogeneity at the interface that could be responsible for magnetic inhomogeneity. Moreover, for the disk sample we measured the dependence of the temperature $T$ on the junction critical current $J_c$ and found it to be proportional to $(1-t)^2$ (with $t = T/T_c$, and $T_c$ the critcal temperature of the proximized device), as was predicted for clean HMF triplet junctions~\cite{Eschrig2003}.

\section{Experimental Methods}\label{sec5}
LSMO (40~nm) was deposited on a (001)-oriented (LaAlO$_3$)$_{0.3}$(Sr$_2$TaAlO$_6$)$_{0.7}$ (LSAT) crystal substrate at 700 $^{o}$C in an off-axis sputtering system with a background pressure of 10$^{-7}$~mbar. The growth pressure was 0.7~mbar with Ar:O (3:2) mixing atmosphere. Information on the characterization of the LSMO films is given in the Supporting Information (SI) Section \textcolor{blue}{I}. After cooling down to room temperature at a rate of 10 $^{o}$C/min, the NbTi layer (60~nm) was deposited on LSMO $in~situ$. The bilayer NbTi/LSMO was pre-patterned to define contacts pad around a central area through ebeam lithography and argon etching.

Next, a gallium focused ion beam (Ga$^{+}$ FIB)  was utilized to structure the bilayer nanopattern, using a beam current of $\sim 30$~pA. We fabricated disk, square, and bar geometries, the latter with two different aspect ratios (length-between-contacts to width) of 3:1 and 5:1. The typical dimension of all structures was 1.3~{\textmu}m. Also by FIB, a trench was made in the middle of the devices with a width (the junction length) of $\sim 20$~nm, For this, a small beam current ($\sim 1.5$~pA) was used. The depth of the trench was defined by the milling time. More details on the device fabrication can be found in the SI Section \textcolor{blue}{I, II}. Shown there are control experiments where the milling time was reduced and increased the milling time to obtain NbTi and LSMO weak links, respectively. \\

Transport measurements were performed in a four-probe configuration; both the electrical transport and magnetoresistance were measured by using a lock-in setup (excitation frequency 77.3~Hz). The resistance-temperature characteristics were taken with a 10~$\mu$A current in a cryostat (IntegraAC, Oxford Instruments). Direction-varying fields and a wide range of temperatures (300~K to 1.5~K) can be achieved in this cryostat. At the setpoint of temperature, the AC current was set to 1~$\mu$A as a background. The current $vs$ voltage measurements were performed by sweeping a DC current superimposed on the background and reading the corresponding voltages, with a field sweep (yielding superconducting quantum interference (SQI) patterns) or a temperature sweep (yielding $I_c(T)$). \\

Micromagnetic simulations were conducted using a GPU-accelerated mumax3~\cite{2014mumax}. The magnetization was set to 5.75$\times10^{5}$~A/m, and exchange stiffness was 1.7$\times10^{-12}$~J/m, yielding an exchange length of $\ell_{ex}$ $\approx$ 2.86~nm~($\ell_{ex} = \sqrt{2A_{ex}/\mu_0 M_s^2}$). According to Ref.~\cite{2011Boschker}, we consider a biaxial anisotropy in the LSMO(40~nm)/LSAT system. Therefore, the constant of biaxial anisotropy was set to 600~J/m$^3$. The damping constant was set to 0.5 artificially to get a high efficiency of convergence. To mimic the real situation, the arms were included in the simulation design. \\

The STEM-EELS data were acquired at 200~kV in a JEOL ARM200cF microscope equipped with a spherical aberration corrector and a Gatan Quantum Dual-EELS spectrometer. The EEL spectra were acquired using the spectrum line mode, with an energy dispersion of 0.25~eV per channel, an acquisition time of 0.5~s and a pixel size of 0.15~nm. Principal Component Analysis (PCA) was performed on the EELS datasets to de-noise the spectra, by using the MSA plug-ins for Gatan DMS analysis toolbox~\cite{watanabe09}.

\section{Results}\label{sec2}

\begin{figure}[ht]
\centering
\captionsetup{width=\textwidth}
\includegraphics[width=\textwidth]{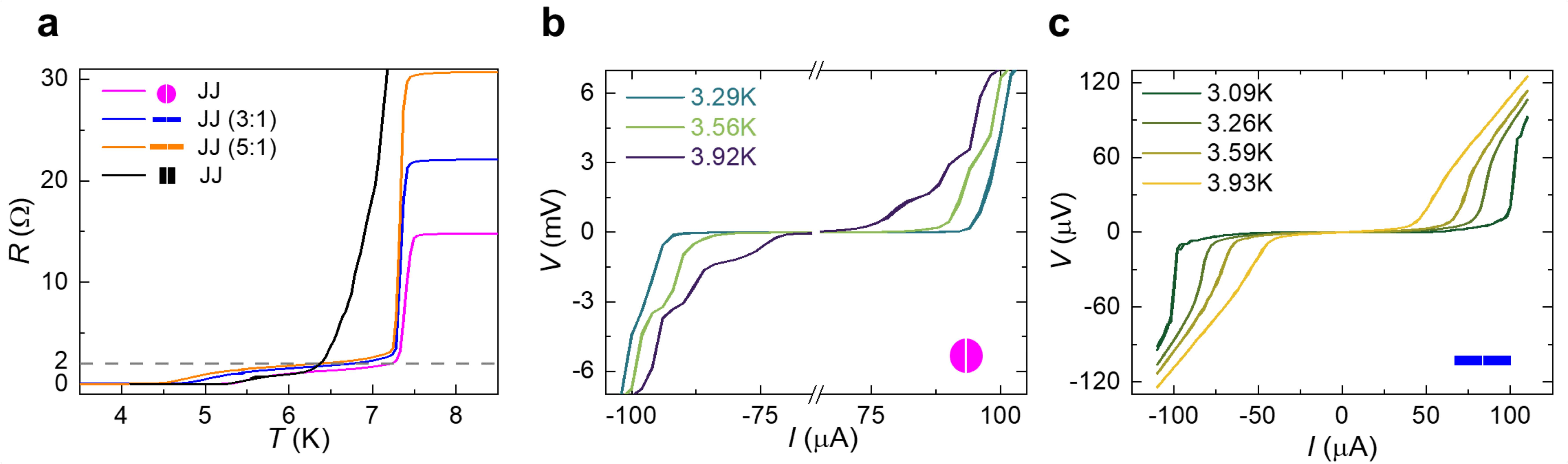}
\caption{Electrical transport. (a) Resistance versus temperature curves for disk-shaped (magenta), square-shaped (black), bar-shaped 3:1 (blue), and 5:1 (orange) devices. The grey dashed line is a guide to the eye at 2~$\Omega$. (b) A plot of current (I) - voltage (V) curves taken at different temperatures on the disk-shaped junction. (c) Measured IV curves on the bar-shaped (3:1) junction at different temperatures. }\label{disk-fig2}
\end{figure}

The resistance $R$ of the differently shaped NbTi/LSMO junctions was measured as a function of temperature $T$, as shown in Fig.\ref{disk-fig2}a. The normal state resistance for all devices is around 20~$\Omega$, except for the square one which is 75~$\Omega$, due to a different contact geometry. The traces show two transitions with decreasing temperature. The first one corresponds to the superconducting transition of NbTi at $\sim$ 7.5~K, and is followed by a plateau, which corresponds to the resistance of the LSMO weak link. As the temperature further decreases, the resistances fully go to zero, with broad tails, indicating that the weak links become superconducting in all junctions, regardless of their geometry and aspect ratio. Since LSMO is a half-metal, the proximity effect must be carried by LRT correlations. In Fig.\ref{disk-fig2} we also show current $I$ versus voltage $V$ (IV-characteristics) for different temperatures of the disk-shaped and the (3:1) bar-shaped junctions, measured by sweeping the current back and forth. \\
%

The data clearly show the onset of voltage at the critical current $I_c$, without visible hysteresis. The multiple steps in the IV curves of the disk junction at higher temperatures result from destroying superconductivity in nanostructured regions in the junction.

As mentioned, in previous work on Nb/Co devices we found proximity supercurrents in disk junctions, resulting from the spin texture of the device. Nb/Co-based bar-shaped junctions were never superconducting due to their uniform magnetization. In the case of LSMO-based junctions, we find a temperature-dependent $I_c$ for all junctions, regardless of the device geometry. For an easy comparison between the Nb/Co and NbTi/LSMO experiments, see Table I at the end of the paper.
\subsection{Spatial supercurrent distribution}\label{subsec3}

\begin{figure}[ht]
\centering
\includegraphics[width=0.93\textwidth]{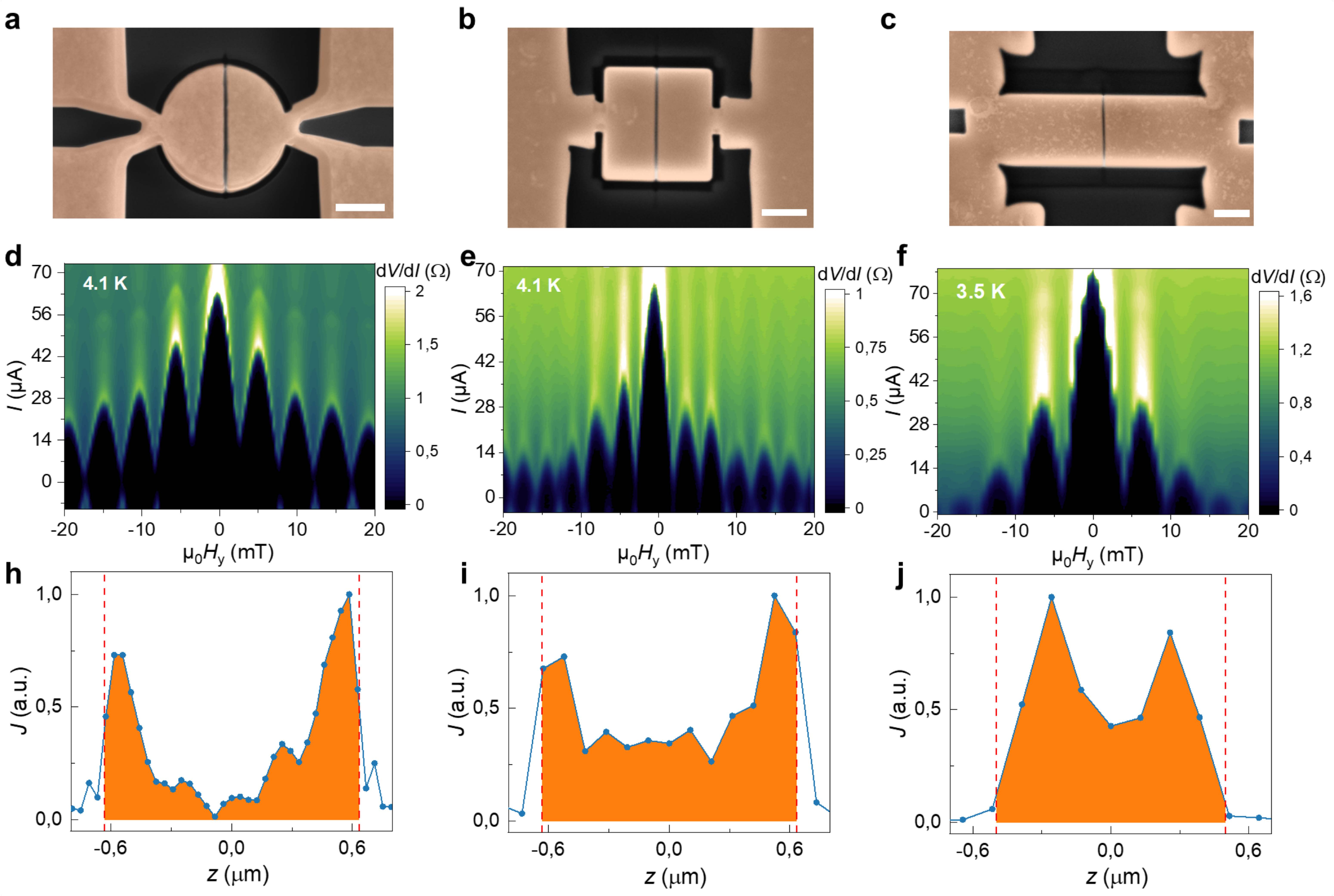}
\captionsetup{width=.93\textwidth}
\caption{Spatial distribution of supercurrents versus geometry. (a-c) Top-view false-colored scanning electron micrographs of disk-shaped, square-shaped, and bar-shaped (3:1) devices. Here, the color indicates the NbTi/LSMO bilayer and all scale bars are equal to 500 nm. (d-f) Measured I$_c$(B$_{\perp}$)-patterns using a $B_{\perp}$ sweep. The color scale gives the differential resistance. (h-j) The results of Fourier analysis. The red-dashed lines indicate the size of the weak link. }\label{disk-fig3}
\end{figure}

To study the relative distribution of critical currents along the junction, we recorded I$_c$(B$_{\perp}$)-patterns (with $B_{\perp}$ a magnetic field perpendicular to the junction plane) of the differently shaped junctions at 4.1~K (disk and square) and 3.5 K (3:1 bar). The results (see Fig.\ref{disk-fig3}d-f) show a clear dependence on the sample geometry. For the disk, the I$_c$(B$_{\perp}$)-pattern is two-channel-like, where the side lobes have a period comparable to the central peak, and the I$_c$(B$_{\perp}$) oscillations decay gradually and less fast than the 1/B$_{\perp}$ behavior expected for a single junction. This points to the existence of two superconducting channels in the disk-shaped NbTi/LSMO junction.
Contrarily, for the bar (aspect ratio 3:1), we find a typical single-junction Fraunhofer-like pattern, the same as the longer bar-shaped (5:1) junction (SI Section II). The square-shaped junction is somewhat in between. We extracted the periodicity for each pattern by using a voltage criterion (see SI, Section \textcolor{blue}{III}). In the case of the disk-shaped junction, the width of the central peak is about 5.91~mT, and that of the first lobe is about 5.04 mT, yielding a ratio of $\sim$ 1.17. For the square-shaped and bar-shaped junctions, we obtain ratios of $\sim$ 1.37 and $\sim$ 1.45, respectively. The difference may be intrinsic in these junctions or correlated with the magnetic structures, as we will discuss later. \\

To convert the I$_c$(B$_{\perp}$)-pattern into the spatial distribution of the critical supercurrent, we employ a Fourier method and reconstruct the supercurrent density along the $z$ axis (along the trench)~\cite{1971Dynes,2017Lahabi,2022Fermin,2023Fermin}. The analysis requires the effective junction length $L_{eff}$, which is often taken as $L_{eff}$ = $d$+2$\lambda_L$. Since the thickness of NbTi electrodes (60 nm) is smaller than their London penetration depth ($\lambda_L$; of order 0.5~{\textmu}m), $L_{eff}$ is rather determined by the lateral geometry of the electrodes. This works out as \( L_{eff} = (1/3.8)*L, (1/2.5)*L, (1/1.842)*L\) for the disk-shaped, square-shaped and bar-shaped junctions, respectively, where $L \approx 1.3~\mu m$ is the junction width~\cite{2023Fermin}. We find the supercurrents to be largely constricted to the rims of the device in the disk-shaped LSMO junction (Fig.\ref{disk-fig3}h). This is similar to the disk-shaped Nb/Co junctions, although the width of the channels is larger (measured by the full width at half maximum $\sim$200~nm). However, as the geometry of the junction changes from disk to square to bar (Fig.\ref{disk-fig3}i and j), the density of supercurrent in the middle regions becomes more and more pronounced (Fig.\ref{disk-fig3}h-j). Notably, the control experiment on disk-shaped MoGe/Ag junction also proves singlet supercurrents are distributed homogeneously across the trench, rather than being localized at the rim~\cite{2022Fermin}.

\subsection{The effect of in-plane fields on the supercurrent distribution}\label{subsec4}

In Nb/Co disk junctions, we found that the proximity effect, and in particular the rim supercurrents, were a consequence of the spin vortex texture. Therefore, the transport characteristics of the device are very sensitive to the exact spin texture of the weak link, for example by moving the vortex core using an in-plane (IP) field.

\begin{figure}[ht]%
\centering
\captionsetup{width=\textwidth}
\includegraphics[width=\textwidth]{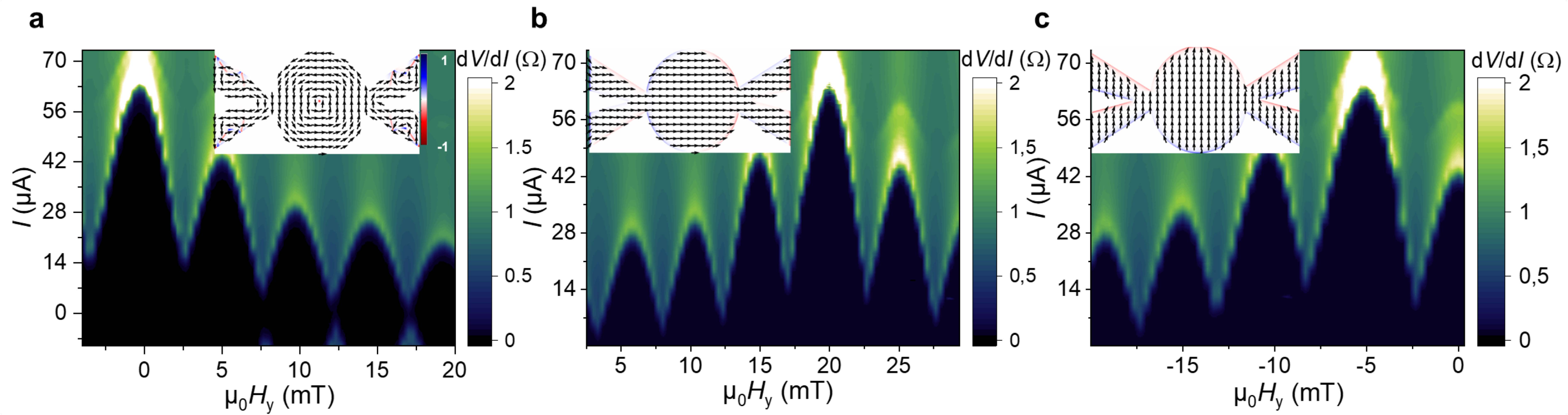}
\caption{I$_c$(B$_{\perp}$)-patterns recorded on a disk-shaped bilayer junction under simultaneous application of IP fields, measured at 4.1~K. (a) I$_c$(B$_{\perp}$)-pattern measured at zero IP field. The micromagnetic simulation in the inset demonstrates the LSMO disk has a ground state of a magnetic vortex. I$_c$(B$_{\perp}$)-patterns in the presence of 200~mT perpendicular (b) or parallel (c) to the trench. Accordingly, the insets display the fully-magnetized state in both cases. The color bar represents $z$ components in the micromagnetic simulation.} \label{disk-fig4}
\end{figure}

In order to explore the similarity, we apply IP fields on the NbTi/LSMO disk junction to change the magnetic ground state, which is also a magnetic vortex (according to micromagnetic simulations; details in Methods). The results, for fields of 200~mT both perpendicular and parallel to the trench, taken at 4.1~K, are shown in Fig.\ref{disk-fig4}. Surprisingly, we find no change in the critical current while applying an IP field. Moreover, the I$_c$(B$_{\perp}$)-patterns persist, both when IP fields are applied perpendicular or parallel to the trench. Even in a 200~mT field, for which micromagnetic simulations demonstrate that the disk is fully magnetized, we find hardly any change in the two-channel I$_c$(B$_{\perp}$)-patterns (Fig.\ref{disk-fig4}b,c). Note that we see small shifts in the central peaks with respect to zero out-of-plane field, but these are due to the misalignment between the sample plane and field direction. We also paid attention to small fields ($\sim$ 10~mT) when small motions of the vortex core are expected. We found no sign of a $0-\pi$ transition as observed in the Nb/Co system. This agrees with the predicted implication that such a transition cannot exist in a HMF~\cite{2008Eschrig}. More details are given in SI Section \textcolor{blue}{III}.

We conclude from our IP field experiments that spin texture is not responsible for the proximity effect in our LSMO-based junctions. Specifically, the observed phenomena are very different from the observations in the Nb/Co disk junctions, where changes in spin texture lead to changes in $I_c$ with fields of mT's, and where $I_c$ is fully quenched when the disk becomes homogeneously magnetized. Here, $I_c(0)$ does not even decrease, emphasizing the robustness of the LRT supercurrents Another possible source of LRT correlations would be a magnetically disordered interface. We come back to this below. Still, the occurrence of the rim currents appears to be coupled to the geometry. In order to test this further, we reshaped a disk junction by cutting off one side. Interestingly, rim supercurrents remain on the curved side, but are absent on the flat side (See SI Section \textcolor{blue}{IV}). Since we ruled out spin texture, we conclude that there is a generator at play at, or close to, the LSMO/NbTi interface. In order to further investigate this, we attempted to fabricate junctions with an altered interface, by inserting a thin Ag layer ($\sim$ 10~nm). However, the Ag turned out to form islands rather than a continuous layer, and the devices showed very similar properties to those without Ag. More details are given in SI Section~\textcolor{blue}{V}.

\subsection{Dependence of I$_c$ on temperature}\label{subsec5}

\begin{figure}[ht]%
\centering
\includegraphics[width=8cm]{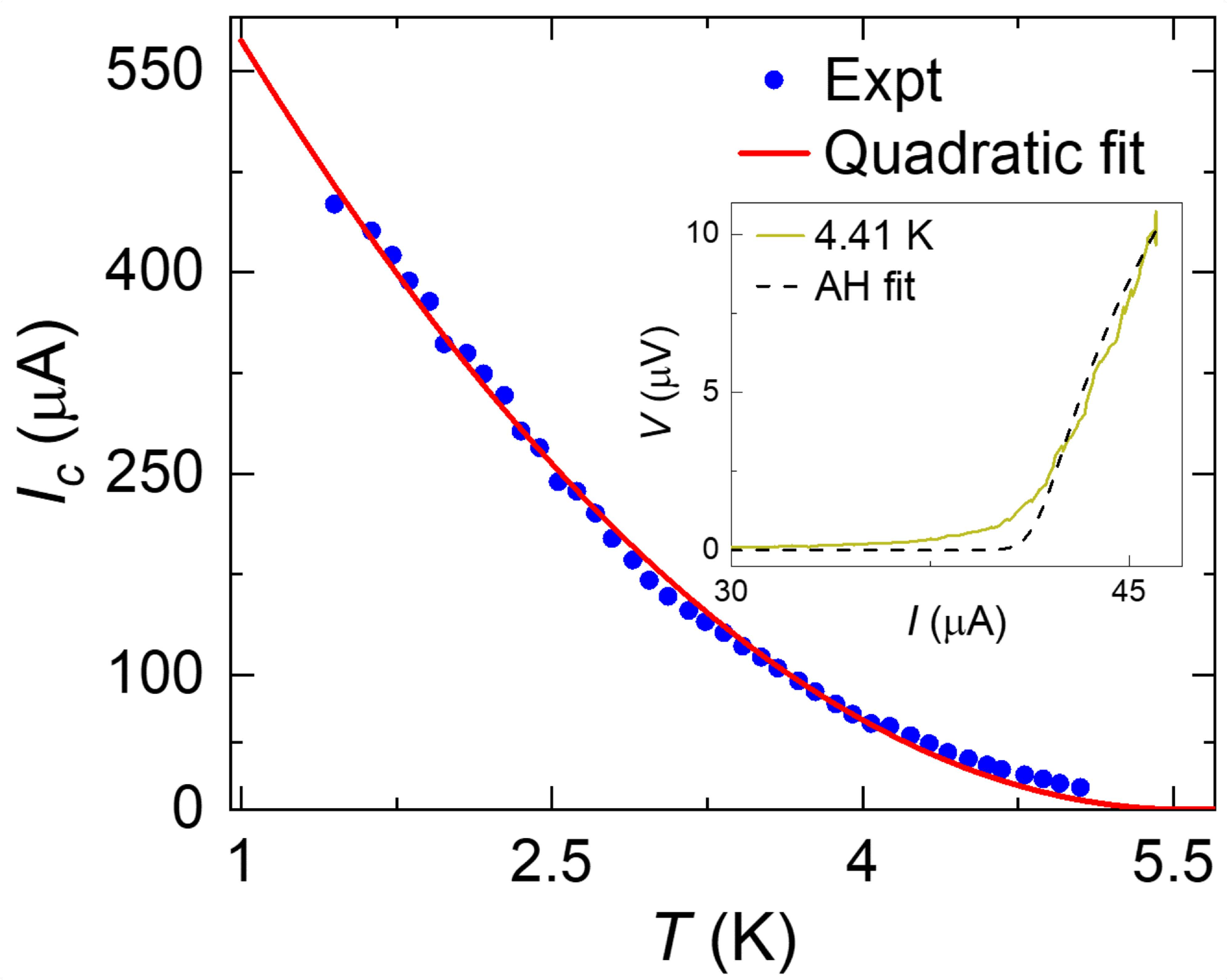}
\caption{$I_c$ as a function of temperature $T$. The inset shows how $I_c$ values are obtained by fitting the IV curve (yellow) to the AH theory (black-dashed line) at 4.41~K, instead of being extracted with a resistance criterion. The extracted values of $I_c$ are plotted by blue dots and well fitted by (1-$t$)$^2$ with $t$ = $T/T_c$ (red line), with $T_c$ = 5.5~K} \label{disk-fig5}
\end{figure}

One of the promises of supercurrents in HMFs is that the density can be very high, and therefore we measured $I_c(T)$ down to 1.5~K. Although $I_c$ is well-defined at low temperatures, near the critical temperature the $IV$ characteristics become rounded around $I_c$, which results in a slight ambiguity in determining $I_c$. The latter is believed to result from phase slips~\cite{2021Blom}.
In this regime, we find $I_c$ by fitting the $IV$ curves to a model proposed by Ambegaokar and Halperin (AH)~\cite{1969AH}. The values for $I_c$ obtained this way are slightly different from the values extracted with a resistance criterion. More details are given in SI Section \textcolor{blue}{VII}, including the relevant $IV$ characteristics. Also, we note that the fitting yields an average value for $R_N$ of 0.8~$\Omega$. This is significantly lower than the plateau in the $R(T)$ curve for this sample, which is around 2~$\Omega$, indicating that the interface resistance is not small, and the interface is not very transparent.
The results for $I_c(T)$ are summarized Fig.\ref{disk-fig5}. They show a parabolic dependence, and a current density $J_c = I_c/(d*L)$ of about $1.7\times10^{10}$~A/m$^2$ at 1.5~K. If we consider that the currents mostly are confined to the rim of the disk, over a distance of $\sim$~200~nm, $J_c$ is $\sim 1.1\times10^{11}$~A/m$^2$. Such high-density spin-polarized supercurrent holds promise for practical applications in superconducting spintronics.\\

Previously, $I_c(T)$ of HMF junctions were analyzed by an analogy to long diffusive SNS junctions~\cite{2010Anwar,2001Dubos}. Here this analogy cannot be made, since the diffusive coherence length is roughly the size of the weak link. To show this, we estimate a mean free path $\ell_H$ of 6.5~nm~\cite{2001Nadgorny} using the measured resistivity of LSMO (40~$\mu\Omega$cm). Next, using a Fermi velocity value $v_F \sim 7.4\times10^5$~m/s this leads to a diffusion constant $D = v_F*\ell_H$/3, and using $T_c$~= 5.5~K, we find the diffusive coherence length $\xi_{F} = \sqrt{\hbar D/(2\pi k_B T_c)}$ to be $\approx$~19~nm, which is roughly the size of the weak link.  The same conclusion follows from considering the Thouless energy $E_{Th} = \hbar D/ L^2$, which is about 2.6~meV, clearly larger than the gap $\Delta$, which is 0.9~meV.\\

A theoretical framework for describing HMF-based junctions, which is more suited than the analogy to diffusive SNS junctions, was given in Refs.~\cite{2003Eschrig,2008Eschrig}. For the clean case, the strength of $I_c$ is determined by the two spin scattering channels $(\uparrow\downarrow - \downarrow\uparrow)$ and $(\uparrow\uparrow)$ from the superconductor to the half-metallic spin-up band. Close to $T_c$, $I_c(T)$ shows a (1-$t$)$^2$ dependence (with $t=T/T_c$), and a maximum at lower $T$, typically around $t=0.2$. The maximum is robust, and is also predicted for long diffusive junctions~\cite{2008Eschrig}, but has never been observed. In the current experiment, the data are well described by (1-$t$)$^2$, using $T_c$ = 5.5~K, see Fig.\ref{disk-fig5}. The experimentally lowest accessible temperature is 1.5~K, or $t = 0.3$, so the issue of the peak in $I_c$ remains open.

\section{STEM inspection of the interface}\label{sec-stem}
\begin{figure}[ht]%
\begin{minipage}{\textwidth}
\captionsetup{width=0.8\textwidth}
\includegraphics[width=0.8\textwidth]{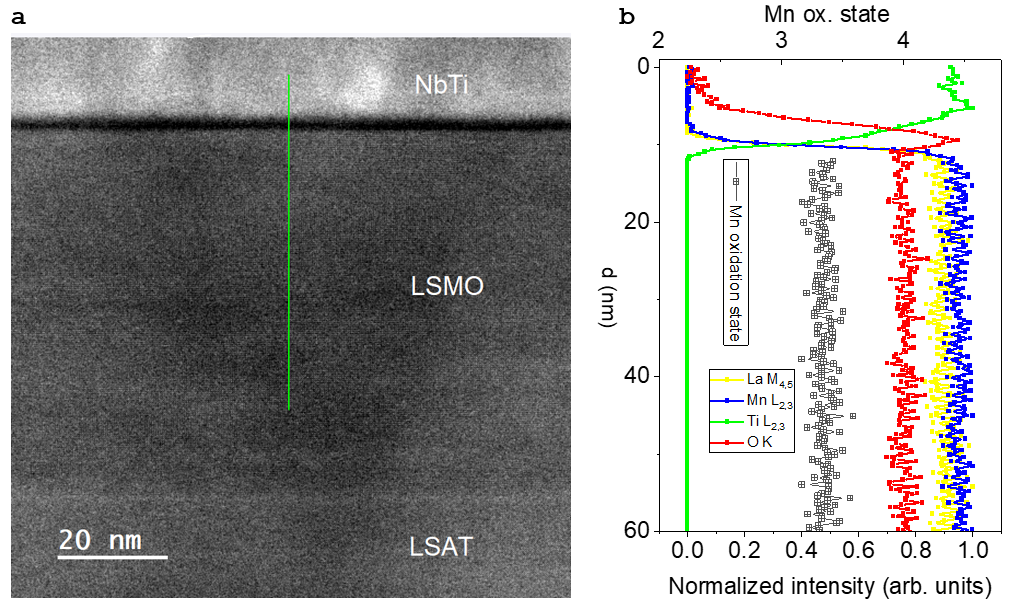}
\caption{(a) High resolution high-angle annular dark field image of the sample interface of LSMO/NbTi deposited on a (001)-oriented LSAT crystal substrate. The green line indicates the zone where an Electron Energy Loss line scan has been performed for the chemical profile shown in (b). (b) Mn oxidation state and chemical profiles of La M4,5, Mn L2,3, O K and Ti L2,3 edges along the NbTi/LSMO interface. A clear peak is visible in the oxygen signal at the interface.} \label{disk-TEM}
\end{minipage}
\end{figure}

An aberration corrected STEM-EELS microscope was used to study the structural and chemical properties of the LSMO/NbTi interface. Fig.\ref{disk-TEM}a displays a high angle annular dark field image showing the high quality LSMO layer and the disordered LSMO/NbTi interface. Fig.\ref{disk-TEM}b shows a chemical profile of the layers from an averaged electron energy loss spectrum image, measured by the normalized intensities at the ionization edges of La M4,5, Mn L2,3, O K, and Ti L2,3. The interface is quite sharp, with no significant intermixing, although a few atomic layers around the interface may well be a mixture of La, Sr, Mn, Nb and Ti. The Nb L2,3 intensity profile vanishes along with the Ti L2,3 (See the averaged EEL chemical map in SI Section \textcolor{blue}{VI}), which proves the mixing along the few atomic layers of the incoherent interface. What stands out is a small peak in the oxygen concentration that we found in two out of three samples. Since the oxygen concentration is all-important for the magnetic state of the manganese, this suggests that the interface can easily be magnetically disordered. In order to get more precise information concerning Mn oxidation state along the LSMO layer, its numerical value was estimated from the L2/L3 intensity ratio of the signals (L23 ratio) in the EEL spectra, according to the procedure described by Varela \textit{et al.}~\cite{varela09} Using this experimental relationship and the  L23 ratios obtained from our spectra, a numerical estimate of Mn oxidation states was obtained.  Fig.\ref{disk-TEM} summarizes the results obtained, indicating that the Mn oxidation state is a Mn$^{3+}$-Mn$^{4+}$ mixture (as expected), and fully constant along the LSMO layer up to the interfacial region of the sample. This reinforces the conclusion that the generator is to be found in the disordered interface.

\section{Discussion}\label{sec-discuss}
Above, we demonstrated the occurrence of a superconducting proximity effect in LSMO-based junctions of various shapes. In this section, we discuss the possible origin of the effect. First note that the superconducting transport is carried by triplet Cooper pairs, as is indicated by the quite pure $(1-t)^2$ dependence of $I_c(T)$. For short junctions with a normal metal interlayer, the expected behavior is $1-t$~\cite{1979Likharev}. A quadratic dependence, or an upward curved behavior of $I_c(T)$, can in principle be found in S/N/S structures~\cite{1964deGennes}, but that would concern long junctions, whereas our junctions are in the short junction limit. We further discuss this point in SI Section \textcolor{blue}{VII}.

Next, we find that the supercurrent generation is independent of the magnetic texture of the HMF. By applying in-plane fields, we observe the supercurrents to persist, even though the spin texture has been quenched. This shows that another generator is at play for the triplets than the vortex magnetization pattern, in line with the fact homogeneously magnetized bar-shaped junctions show equally strong supercurrents, and confirming our results on long junctions~\cite{2024Yao}. The inescapable conclusion is that the LRT correlations are generated at the NbTi/LSMO interface. A possible source of LRT correlations would be inhomogeneous interfacial magnetism, stemming from the atomically disordered interface. Disorder would generally favor antiferromagnetic interactions, which are less sensitive to magnetic fields. \\

Finally, we discuss the relation of the LRT generator to the highly confined triplet supercurrents at the rims of the disk-shaped devices. In the Nb/Co disk junctions studied before, these rim supercurrents were believed to be a direct result of the triplet nature of the supercurrents. Specifically, in the Nb/Co system the generator was identified as resulting from the synergistic effects of an effective spin-orbit interaction (provided by the vortex magnetization) and sample boundaries. This entails that LRT transport in the Nb/Co system can only emerge at the rims of the device, and consequently, rim supercurrents are very sensitive to in-plane fields. Contrarily, in the NbTi/LSMO system (for a comparison, see Table I), triplet generation is independent of spin texture. In the LSMO-based junctions, we observe rim supercurrents in the disk that become weakened in the square and bar. We suspect that the difference between the Nb/Co and NbTi/LSMO systems is connected to the high spin polarization of the LSMO, where only one spin state can exist, but also this remains an open question.

\begin{table}[H]
\centering
\begin{tabular}{|c |>{\centering}p{1cm} | >{\centering}p{1cm} |
>{\centering\arraybackslash}p{1.5cm} |} \hline
        & \multicolumn{3}{c|}{SQI patterns  }  \\ \hline
           & disk    & bar  & IP field \\ \hline
Nb/Co      &  S      & not sc  &  yes    \\ \hline
NbTi/LSMO  &  S      & F  &  no  \\ \hline
\end{tabular}
\caption{\raggedright Comparison of SQI patterns for disks and bars of Nb/Co and NbTi/LSMO. S means a SQUID pattern, F means a Fraunhofer pattern, not sc means the junction was not superconducting. The column \linebreak IP field shows whether the patterns \textit{for the disks} \linebreak change under  an in-plane magnetic field.}
\end{table}

\section{Conclusion}\label{sec4}
In summary, we have unambiguously shown supercurrents and Josephson coupling in lateral NbTi/LSMO junctions with different geometries, and examined their origin. Surprisingly, these currents remain robust against large in-plane fields that are able to erase the spin texture, indicating that the spin texture is not relevant for producing such currents. Instead, intrinsic magnetic inhomogeneity in the LSMO top layer may be the key, which would also explain triplet generation in the LSMO/YBCO system. By performing Fourier analysis on the various I$_c$(B$_{\perp}$)-patterns, we find rim supercurrents in the disk-shaped junctions. The origin of these rim supercurrents is tied to the combination of triplet transport and disk-shaped junction geometry. However, their exact origin is currently left unexplained. On a final note, the robust triplet generator that we find is helpful for researching further fundamental questions involving spin-polarized supercurrents, but less so for applications, because it lacks \textit{control}. For that, a different interface geometry will have to be devised.

\noindent


\section{Supporting information}
Supporting Information is available free of charge at https://xxxx. It gives more detailed descriptions of the following: a characterization of the epitaxy of LSMO films (Section I); results on junctions with shallow and deep trenches (section II); critical currents extracted with a resistance criterion, and SQI patterns at small in-plane field (Section III); a disk-shaped NbTi/LSMO junction with a flat side (Section IV); measurements and TEM data on a LSMO/Ag/NbTi device (Section V); an average EEL chemical map of the LSMO/NbTi interface (section VI), and the determination of $I_c$ and a calculation of $E_{Th}$ for long diffusive junctions (Section VII).

\section{Author information}
All authors contributed to designing the experiment. YJ and RF fabricated the devices, YJ performed the measurements and the data analysis, with input from RF, JA and KL. MCP performed the TEM analysis. YJ and RF wrote the manuscript, with input from all authors. \\

\section{Acknowledgments}
This work was supported by the project ‘Spin texture Josephson junctions’ (project number 680-91-128) and by the Frontiers of Nanoscience (NanoFront) program, which are both (partly)  financed by the Dutch Research Council (NWO). YJ is funded by the China Scholarship Council (No. 201808440424). The work was further supported by EU Cost actions CA16218 (NANOCOHYBRI) and CA21144 (SUPERQMAP). It benefited from access to the Netherlands Centre for Electron Nanoscopy (NeCEN) at Leiden University. MC received support from the Spanish state research agency AEI through grants PID2020-118078RB-I00 and from the Regional Government of Madrid CAM through SINERGICO project Y2020/NMT-6661 CAIRO-CM.  Electron microscopy observations were carried out at ICTS-CNME in UCM. The authors acknowledge the ICTS-CNME for offering access to their instruments.


\bibliography{LSMOdisk}
%



\end{document}